\documentclass[preprint,aps,showpacs]{revtex4}

\usepackage{ams,amsmath,color}
\usepackage[ansinew]{inputenc} 
 \usepackage{verbatim}

\begin{document}


\title{Energy transfer processes in Er-doped SiO$_2$ sensitized with Si nanocrystals}

\author{I. Izeddin, D. Timmerman, and T. Gregorkiewicz}
\affiliation{Van der Waals-Zeeman Institute, University of Amsterdam
\\ Valckenierstraat 65, NL-1018XE Amsterdam, The Netherlands}

\author{A.S. Moskalenko}
\affiliation{Institut f\"ur Physik, Martin-Luther-Universit\"at Halle-Wittenberg,
Nanotechnikum-Weinberg, Heinrich-Damerow-St. 4, 06120 Halle, Germany}
\author{A.A. Prokofiev, I.N. Yassievich}
\affiliation{Ioffe Physico-Technical Institute, Russian Academy of Sciences \\
Politekhnicheskaya 26, 194021 St. Petersburg, Russia}

\author{M. Fujii}
\affiliation{Department of Electrical and Electronic Engineering, Faculty of Engineering, Kobe University, Rokkodai, Nada, Kobe 657-8501, Japan}

\date{\today}

\begin{abstract}
We present a high-resolution photoluminescence study of Er-doped
SiO$_{2}$ sensitized with Si nanocrystals (Si NCs).
 Emission bands originating from recombination of excitons confined in Si NCs, internal transitions within the 4f-electron core of Er$^{3+}$
 ions, and  a band centered at $\lambda\approx$1200 nm have been identified. Their kinetics have been investigated in detail.
 Based on these measurements, we present a comprehensive model for
energy transfer mechanisms responsible for light generation in
this system.
  A unique picture of energy flow between the two subsystems is developed, yielding truly microscopic information on
 the sensitization effect and its limitations. In particular, we show that most of the Er$^{3+}$ ions available in
  the system are participating in the energy exchange. The long standing problem of apparent loss of optical activity of
  majority of Er dopants upon sensitization with Si NCs is clarified and assigned to appearance of a very efficient energy
  exchange mechanism between Si NCs and Er$^{3+}$ ions. Application potential of SiO$_{2}$:Er sensitized by Si NCs is discussed
   in view of the newly acquired microscopic insight.
\end{abstract}


\maketitle
\section{Introduction}\label{subsec:chap2_sec2_subsec1}
SiO$_2$ matrix doped with Er$^{3+}$ ions and Si nanocrystals (Si NCs) is intensively investigated as
an interesting system where room temperature 1.5 $\mu$m Er-related emission can efficiently be induced
 by non-resonant excitation; some time ago it has been noted that photo- and electroluminescence of Er$^{3+}$ ions
 in SiO$_2$ can be effectively sensitized with Si quantum dots \cite{Kenyon_JPM1994, Fujii_APL1997, Chryssou_Kenyon_APL1999, Pacifici_PRB2003, Kik_Polman_APL2000}.
 In photoluminescence, upon illumination, incoming photons are predominantly absorbed by band-to-band transitions in Si NCs.
  Since the indirect band structure of Si is preserved also in the nanocrystalline form \cite{Kovalev_PRL1998}, electron-hole
   pairs generated in this way are characterized by a relatively long lifetime. This enables energy transfer to Er$^{3+}$ ions
   located in vicinity of Si NCs. In that way a channel of indirect excitation of Er dispersed in SiO$_2$ is created and
    the 1.5 $\mu$m Er-related emission appears. Its temporal characteristics comprises a microsecond rise time,
     corresponding to Si NC-to-Er energy transfer, followed by predominantly radiative and temperature independent decay
      in the millisecond range, characteristic for Er$^{3+}$ ions in SiO$_{2}$. It has been concluded
      that dispersion of Si nanocrystals  in Er-doped SiO$_2$ matrix SiO$_2$:(Er, Si NCs) combines to a certain extent
       positive features of Er-doped crystalline Si with those of Er-doped SiO$_2$:

\begin{itemize}
\item
In contrast to the situation for SiO$_2$:Er, introduction of Si NCs enables indirect excitation of Er$^{3+}$  ions.
 This process is non-resonant and relatively efficient, with an (effective)
 excitation cross-section of $\sigma \approx 10^{-17}-10^{-16}$ cm$^{-2}$, which
 represents an increase by a factor $10^3$ \cite{Pacifici_PRB2003} in comparison to SiO$_{2}$:Er.

\item In contrast to the situation in crystalline Si:Er, emission
from Er$^{3+}$ ions does not suffer from thermal quenching and is readily observed at room
temperature, similar to SiO$_{2}$:Er \cite{Kik_Polman_APL2000, Pacifici_PRB2003}.
\end{itemize}

These promising characteristics rose considerable hopes on possible applications of the SiO$_2$:(Er, Si NCs) for Si photonics
 and specific devices have been proposed \cite{Castagna_Coffa_PE2003, Nazarov_Skorupa_APL2005, Walters_Atwater_NatMat2005}.
 Particularly  attractive is the prospect application of SiO$_2$:(Er, Si NCs) for development of a flash-lamp pumped optical
 amplifier---a much welcome replacement for the currently used fiber amplifier which requires resonant and high power laser
 pumps for its operation. In order to achieve that, the Si NC-induced sensitization process of Er emission in SiO$_2$
  has to be thoroughly understood.

In the past, F\"{o}rster (dipole-dipole) mechanism \cite{Foerster} has been proposed in order to
explain the energy transfer from Si NCs to Er$^{3+}$ \cite{Dood_PRB2005, Agranovich_1982} and
different locations of Er in respect to Si NC have been considered \cite{Wu_PRL2001,
Senter_Coffer_PRL2004}. In addition to the relatively slow ($\mu$s range) NC-mediated Er$^{3+}$
excitation, also a much faster (100 ns range) and usually dominating process has been concluded
\cite{Fuji_JAP2004}. Its physical origin has been considered recently \cite{Savchyn_PRB2007} and
participation of special luminescence centers in erbium excitation has been postulated in
particular for low temperature annealed samples.
 However, in spite of a considerable progress by both
modelling and experiment
 \cite{Pacifici_PRB2003, Imakita_Fujii_PRB2005}, many issues still lack sufficient explanation. In this field, while
 reports on optical gain have been published \cite{Lee_LT2005}, detailed investigations revealed important
  drawbacks of the SiO$_2$:(Er, Si NCs) system: it has been realized that (i) only a relatively small part of all
  the Er$^{3+}$ ions is susceptible to the indirect excitation via Si NCs \cite{Wojdak_PRB2004, Kik_Polman_JAP2000},
  and (ii) upon introduction of Si NCs, a considerable portion of Er dopants loses so-called ``optical activity''and does not contribute photons regardless of
   the excitation mode (via Si NCs or direct by resonant pumping) \cite{Wojdak_PRB2004}. Therefore it has become clear
   that dispersion of Si NCs in SiO$_2$:Er is a challenging and complex physical system, which has to be understood, and
    possibly engineered at a microscopic level, before device applications can be considered.

 In our previous contribution \cite{Izeddin_PRL2006}, sub-microsecond Er-related luminescence from the
SiO$_{2}$:(Er, Si NCs) system was reported, and an Auger-facilitated energy transfer process
between carriers within the quantized levels of the NCs and Er$^{3+}$ was proposed as the
excitation and de-excitation mechanism. It was also shown that up to $\sim$ 50\% of the total Er
content is involved in this process and contributes to the sub-$\mu$s emission. In this paper, we
study in depth temporal details of emission bands from SiO$_2$:(Er, Si NCs). Using optical
excitation with nanosecond pulses and time-correlated photon counting detection mode, we resolve
true kinetics of
 emissions related to Si NCs and Er$^{3+}$ ions present in the investigated material. Based on this information, we propose a complete
 microscopic scenario of energy transfer processes in the SiO$_2$:(Er, Si NCs) system. The chosen approach allows us to identify emissions
 from most of the Er dopants available in the material. Using a theoretical model, we discuss microscopic aspects of energy transfer
 processes between Si NCs and Er$^{3+}$ ions in relation to their (mutual) location in the SiO$_2$ matrix.

\section{Experimental}
For the purpose of this study, a series of Si NC- and Er-doped SiO$_{2}$ 2 $\mu$m layers were
prepared by radio frequency co-sputtering on a SiO$_{2}$ substrate. The samples were characterized
by different concentration of Er dopants and size and concentration of Si NCs. Details of sample
preparation procedure can be found in Ref.~\cite{Fujii_APL1997}. Following preliminary selection,
the detailed investigation of recombination dynamics have been conducted on a particular sample
with the most intense 1.5 $\mu$m Er-related emission. It was characterized  by atomic
concentrations of 0.11\% of Er, 41.8\% of Si, and 12.6\% of excess Si. After the sputtering
procedure, the sample underwent annealing at 1100 $^{\circ}$C which resulted in formation of Si NCs
with the average diameter of 3.1 nm with a size dispersion of $\approx$ 14\%, and a density of [NC]
$\approx$ 4.1$\times$10$^{18}$ cm$^{-3}$. A twin sample without Er dopant and the same
concentration and size of Si NCs was used for the absorption measurements.

The photoluminescence (PL) experiments were performed under pulsed excitation in
the visible provided by a tuneable optical parametric oscillator (OPO) pumped by
 the third harmonic of a Nd:YAG laser, with pulse duration of 5 ns and repetition rate of 10 Hz.
 In the UV range experiments, the excitation was provided by the third harmonic of a Nd:YAG.
 The samples were placed on a cold finger of closed-cycle cryostat and the measurements were taken at 10 K and room temperature (RT).
  PL spectra were resolved with a TRIAX 320 spectrometer and detected with an InGaAs photomultiplier tube (PMT) with flat response
  in the visible to near-infrared range or a Ge photodiode (Edinburgh Instruments EI-A) connected to a digital oscilloscope
  where signal integration was done. For time-resolved measurements of PL dynamics, the PMT was working in time-correlated
   photon counting mode with temporal resolution up to 2 ns.

The absorption measurements were performed in two experimental configurations: Direct transmission
measurements under OPO pulsed illumination, and in a UV absorption photo-spectrometer.

\pagebreak
\section{Results}\label{subsec:chap2_sec2_subsec3}

\subsection{Photoluminescence spectra}\label{subsubsec:chap2_sec2_subsec3_subsubsec1}
Figure~\ref{fig:Chap2_Sec2_Fig1} shows the PL spectra of the investigated sample at RT and at 10 K.
The measurements have been performed under pulsed excitation at $\lambda$ = 450 nm (2.75 eV),
\textit{i.e.}
 not in resonance with any internal transition of Er. The total time-integrated PL response of the sample is given,
  which reflects the number of photons emitted. Both, Er- and Si NC- related PL bands are observed simultaneously,
   with their mutual intensity ratio determined by the Er concentration, in agreement with earlier studies \cite{Fujii_APL1997}.
   In addition, at low temperature, a weaker band centered at $\lambda$ $\approx$ 1200 nm can be seen;
   it has a short decay time constant and therefore its time-integrated intensity is small compared to the other two bands.

We note that the Er-related emission band broadens at RT, keeping its wavelength integrated
intensity practically constant, also in concordance with previous reports. The broadening at RT is
produced mainly in the higher energy
 side of the spectrum, and the full width at half maximum (FWHM) increases from $\approx$ 6.7 meV to $\approx$ 20 meV
 at 10 K and RT, respectively. This could be due to population of upper states of the multiplet $^{4}$I$_{13/2}$ by thermalization.

The Si NCs show a broad excitonic-related PL band. If we take into
account the NCs size (radius R = 1.55 nm) and
 dispersion ($\approx$ 14\%), the position and width of the excitonic-related PL band can be estimated on the basis
 of the calculated band gap of the NCs as a function of their size \cite{Moskalenko_PRB2007}. The emission energy is given by E = E$_{g}^{Si}$ + E$_{1e}$ + E$_{1h}$ - E{$_{excit}$} - $\hbar \omega \approx$ 1.5 eV
 (with E$_{g}^{Si}$ corresponding to the energy band gap of crystalline silicon, E$_{1e}$, E$_{1h}$
 and E$_{excit}$
  the electron and hole levels quantization corrections and the excitonic correction,  respectively---taken
  from reference \cite{Moskalenko_PRB2007}---, and the phonon energy $\hbar \omega$ = 60 meV), \textit{i.e.} a band centered at 850 nm, ranging
   from $\approx$ 820 nm to $\approx$ 870 nm taking into account their size dispersion. Comparison with experiment shows that the
   center of the band is in very good agreement with the expected value, but the experimentally recorded band is broader than
   expected from the calculations, with the higher energy side of the band suffering from strong temperature quenching.
   Since optical excitation was provided by short wavelength photons of 450 nm (2.75 eV), thus creating hot carriers in the higher
 energy states of the Si NCs, then it is reasonable to consider the origin of the higher energy side of the NC-related PL band
    as arising from ``hot" carriers in the upper electron (and/or hole) levels of the Si NCs.

Superimposed on the Si NCs PL band recorded at 10 K, peaks with positions at 1.26 eV and 1.57 eV
are observed. These correspond to emissions from the second and the third excited states of Er and
disappear at room temperature. We note that in view of
 the exclusively non-resonant excitation mode, the observation of these peaks indicates that the Er$^{3+}$ excitation proceeds,
 at least partly, via higher excited states.

The effective lifetime of the excited state of Er in SiO$_{2}$ with Si NCs lies within 2-3 ms.
Decay of excitons in Si NCs is governed by a stretched exponential function, with a long final
tail, with decay time constant of 20 $\mu$s to 50 $\mu$s as will be shown later. The decay of the
band centered at $\lambda\approx$ 1200 nm is clearly faster, being predominantly non-radiative; the
lifetime shortens further at a higher temperature, leading to a strong reduction of the
time-integrated PL intensity of this band at RT.

\subsection{High-resolution photoluminescence kinetics}\label{subsubsec:chap2_sec2_subsec3_subsubsec2}

Using the time-correlated photon-counting technique to resolve the time evolution of the PL
transient revealed new insight into excitation and de-excitation dynamics in the SiO$_{2}$:(Er, Si
NCs) system \cite{Izeddin_PRL2006}. This experimental technique allows for simultaneous recording
of very different dynamical ranges without suffering from signal distortion. Panel {\bf a} in
Fig.~\ref{fig:Chap2_Sec2_Fig2} presents a time-resolved PL spectrum of the excitonic PL band from
Si NCs. The contour plot shows how the peak of maximum intensity drifts to longer
wavelengths---\textit{i.e.} lower energy---at longer delays. Consequently, the PL decay constant of
the luminescence increases for longer wavelengths. This is understandable when we consider the NCs
size distribution: excitonic PL dynamics in smaller NCs is faster than in larger quantum dots. In
the Si NCs under consideration the phonon-assisted radiative transition dominates over the direct
radiative transition. The probability of such a transition increases rapidly with the decrease of
the NCs size (see Ref.~\cite{Moskalenko_PRB2007}). In the panel {\bf b} of the figure, the
time-resolved PL spectrum of Si NCs is shown in the sub-microseconds time range; we can observe the
center of the band at $\lambda$ = 775 nm (1.6 eV), in contrast to the previous figure. If we
compare the spectra measured at 100 ns and at 100 $\mu$s after the excitation pulse, a difference
of $\approx$ 200 meV separates the centers of the two bands. We can assign the sub-microsecond PL
to recombination of carriers from the higher Si NC states.

Figure~\ref{fig:Chap2_Sec2_Fig3} shows measurements related to the broad PL band appearing in the
spectral range between the NC- and Er-related emissions. In the inset, the spectrum recorded at 200
ns after the excitation pulse and at T = 10 K, shows the broad band centered at $\lambda$ $\approx$
1200 nm. The lifetime of this PL band, as mentioned before, shortens at higher temperatures,
becoming practically negligible at RT. This can be seen in the main panel, where the RT PL
transient kinetics recorded at the maximum of the band (1200 nm) is compared to the PL at 1535 nm.
One can see that after 100 ns the decay of the PL at 1535 nm can be attributed to the Er-related
PL. The origin of the band centered at $\lambda$ $\approx$ 1200 nm is usually identified as
recombination at defects, but its broadness and fast kinetics seem to contradict this
identification; future investigations---out of the scope of this paper---will elucidate this point.

In Fig.~\ref{fig:Chap2_Sec2_Fig4}, the high-resolution kinetics of the excitonic luminescence from
Si NCs can be followed in the panel {\bf a}. An intense PL signal with a fast decay, also
characterized in the sub-$\mu$s time domain, is observed after the excitation pulse. This is
stretched until the final slower tale---of the order of 30 to 50 $\mu$s---is achieved, as shown in
Fig.~\ref{fig:Chap2_Sec2_Fig2}. Panel {\bf b} of the same figure shows the Er-related PL kinetics
for the first microsecond after the excitation pulse, at T = 10 K and RT.

In our previous contribution \cite{Izeddin_PRL2006} we have analyzed detailed kinetics of
Er-related PL sensitized by Si-NC and have shown that it exhibits 3 separate regimes:\\ Regime I -
for t $\leq$ 1 $\mu$s: a strong emission appears immediately following the laser pulse and then
rapidly decays towards a temporary minimum. \\ Regime II - for  1 $\mu$s  $\leq$ t $\leq$  10
$\mu$s: Er-related PL intensity rises slowly again and reaches a  broad maximum whose amplitude is
at least an order of magnitude smaller that the initial value in Regime I. \\Regime III - for t $>$
10 $\mu$s: PL intensity slowly decays with a millisecond time constant. We have also shown that the
microsecond rise of the Er-related PL in Regime II ($\tau^{rise}_{Er} \approx$ 3.5 $\mu$s) is
paralleled by a decay of Si NC PL, which can be characterized by a stretched exponent with
$\tau_{NC} \approx$ 1.2 $\mu$s, illustrating in this way the relatively slow energy transfer
between Si-NCs and Er$^{3+}$ ions. This decay of NC-related PL slows down once Er emission attains
a maximum. At this stage the NC-mediated excitation of Er is completed or saturated
\cite{Wojdak_PRB2004} and therefore further decay of the NC-related PL is decoupled from Er.
Careful inspection revealed that the local minimum of Er PL intensity from which the ``slow'' rise
begins, is equal to approximately half of the maximum intensity attained in Regime II. We recall
that such an initial amplitude of Er PL attained very shortly after excitation - Regime I - has
been reported also before \cite{Kenyon_JAP2004}. Based on the total evaluation of optically active
Er fraction, we conclude that the``residual concentration'' of Er$^{3+}$ ions involved in the slow
excitation/slow recombination process - Regimes II and III - amounts to $\sim$ 0.2\% of the total
Er content.

To validate the idea of hot carriers in the Si NCs being responsible of the sub-microsecond
Er-related luminescence---postulated in our previous contribution \cite{Izeddin_PRL2006}---, we
investigate the intensity ratio between the fast and the slow components of Er PL for different
excitation wavelengths and fluxes. Both, fast (sub-microsecond) and slow (milliseconds) components
of the Er-related transient PL must be governed by different excitation and de-excitation
mechanisms. The intensity ratio of these components is thus likely to depend on the excitation
conditions. Figure~\ref{fig:Chap2_Sec2_Fig5} shows full, high-resolution kinetics of Er$^{3+}$ PL,
under different excitation wavelengths and powers, normalized for the maximum intensity of the
\textit{slow} component. One can observe a higher \textit{fast}-to-\textit{slow} intensity ratio
when higher power or larger photon energy quantum are used for excitation.

\subsection{Excitation cross section measurements}\label{subsubsec:chap2_sec2_subsec3_subsubsec3}

The excitation cross section of Si NCs and Er PL,  $\sigma_{NC}$ and $\sigma_{Er}$, respectively,
is of crucial importance to understand the excitation processes and energy transfer between Si NCs
and Er$^{3+}$ ions. In order to gain insight into this aspect, flux dependencies of Si NCs- and
Er-related PL signals were recorded; a reference sample without Er doping was used for the
NC-related PL measurements. In Fig.~\ref{fig:Chap2_Sec2_Fig6}, PL flux dependencies for different
excitation wavelengths are shown, recorded at 1535 nm and 912 nm, for the Er-doped and Er-free
samples, respectively. The curves have been fitted to the excitation dependence of PL intensity,
derived from the rate equations for pulsed excitation,
\begin{equation} \label{eq:chap2_sec2_eq1}
 I_{PL} \propto N^{*} = N (1 - e^{- \sigma \phi \Delta t})
\end{equation}
where $I_{PL}$ is the time-integrated PL intensity, in arbitrary units; $N^{*}$ is the number
(concentration) of excited emitters (Er$^{3+}$ ions or Si NCs); $N$ is the total number
(concentration) of excitable Er$^{3+}$ ions or NCs; $\sigma$ is the effective excitation cross
section; $\phi$ is the photon flux of excitation; and $\Delta t$ the laser pulse duration. Remark
that formula (\ref{eq:chap2_sec2_eq1}) is valid only in a limit of one radiative exciton per Si NC
\cite{Wojdak_PRB2004}. The effective excitation cross section $\sigma$ is determined from the
experimentally measured dependencies depicted in Fig.~\ref{fig:Chap2_Sec2_Fig6} by fitting them
using formula (\ref{eq:chap2_sec2_eq1}). The resulting values, for both Si NCs and Er luminescence
($\sigma_{NC}$ and $\sigma_{Er}$), are shown in Fig.~\ref{fig:Chap2_Sec2_Fig7} as a function of the
excitation wavelength.

Directly related to the effective PL excitation cross section $\sigma_{PL}$, we studied the optical
absorption coefficient $\alpha$ in our SiO$_{2}$:Si NCs reference sample. The effective (Er- and
NCs-related) PL excitation cross section values (Fig.~\ref{fig:Chap2_Sec2_Fig7}) are plotted in
Fig.~\ref{fig:Chap2_Sec2_Fig8} as a function of the measured Si NCs' absorption coefficient
$\alpha$, for each given excitation wavelength $\lambda_{exc}$. A linear relation between
absorption and excitation can be observed in all the investigated range for Si NCs; in the case of
Er-related luminescence, a change of linear relation is observed at an energy threshold, at $\sim$
2.6 - 2.7 eV ($\approx$ 460 - 470 nm), energy above which the excitation cross section grows faster
than the absorption.

\pagebreak
\section{Theory}\label{subsec:chap2_sec2_subsec4}

In order to rationalize the experimental results gathered in this
study, we consider the possible processes of energy exchange
between carriers confined in Si NCs and Er$^{3+}$ ions in SiO$_2$
(outside NC).

Energy transfer between electrons and holes in NCs and $f$-electrons of Er$^{3+}$ ion is
implemented as an Auger process (\textit{i.e.} via the Coulomb interaction).The energy can be
transferred to an erbium ion either when a confined  electron-hole pair recombines, or when an
intra-band transition of confined carrier occurs. Both processes can be accompanied by multiphonon
transitions to fulfill the energy conservation law, as the energy spectra of both electron systems
are discrete. The momentum conservation law plays an important role in the Auger recombination
processes as the large momentum should be transferred by the electron-hole recombination due to the
indirect band structure of silicon.

Confined electrons and holes energy levels as well as their wave functions calculated in multiband
effective mass approximation \cite{Moskalenko_PRB2007} are used in this consideration. Luttinger
Hamiltonian in the spherical approximation has been used for holes and the strong anisotropy of the
electron effective mass in silicon has been taken into account. The wave function and flux
continuities were used as the boundary conditions \cite{Schenk_JAP1997, Chelikowsky_PRB1977}. When
calculating wave functions outside the NC, isotropic effective masses being equal to $m_0$ and
$5m_0$ were used for electron and hole states correspondingly.  Spin-orbit splitting was neglected
in both Si and SiO$_2$.

The conduction band of Si has six equivalent minima in the first
Brillouin zone, situated in the neighborhoods of the six
$X$-points. The wave function of electron can be presented in the
form:
\begin{equation}
    \psi^{e}_{\nu}=\xi^{e}(\mathbf{r})u_{c\nu}\exp(i\mathbf{k}_{0\nu}\mathbf{r}),
\label{electron1}
\end{equation}
where $u_{c\nu}$ and $\mathbf{k}_{0\nu}$ ($k_{0\nu}=0.85\times2\pi/a_{\mathrm{lat}}$) are the Bloch
amplitude and wave vector corresponding to the bottom of valley $\nu$ ($a_{\mathrm{lat}}=0.54$~nm
is the lattice constant of silicon). Envelope functions $\xi^{e}(\mathbf{r})$ are found as a result
of a numerical solution to the Schr\"{o}dinger equation after separating the angular part
$\exp(im\phi)$ ($m$ can be any integer number) as there is a strong anisotropy of the electron
effective mass: $m_{\|}=0.916m_{0}$, $m_{\perp}=0.19m_{0}$. The electron states are sixfold
degenerate for m=0 and 12 times degenerate for $|m|>0$, as two opposite values $m=\pm|m|$
correspond to the same energy. (This degeneracy is given without taking in account an additional
spin degeneracy). So the states are marked with $Ne_{|m|}$ where the letter $e$ shows that it is an
electron state, $N$ is the main quantum number which shows the order of the energy level for given
$|m|$.  For example, the ground state is marked as $1e_0$, which means that this is the first
electron state with $m=0$.

The approximation used in \cite{Moskalenko_PRB2007} leads to two types of hole states in bulk Si
which correspond to twofold degenerate heavy hole band having mass $m_{h}=0.44m_{0}$ and
nondegenerate light hole band having mass $2m_{h}m_{l}/(3m_{h}-m_{l})=0.12m_{o}$ ($m_{l}=0.16m_{0}$
for Si) \cite{Abakumov_book_1991}. The quantum confinement gives rise to mixing of the states.
There are three types of hole states in spherical quantum dots: i) mixed states ($hm$) formed by
the combination of heavy and light ones, ii) heavy hole ($hh$) states, and iii) light hole ($hl$)
states. Each state is also characterized by the full angular momentum $F$ (0 for the light hole
states and positive integer for the other ones) and it is $2F+1$ times degenerate as the projection
$M$ of momentum $F$ onto the quantization axis (arbitrary selected) can be any integer number
having absolute value not larger than $F$. The space quantization forms a series of each type of
the states with fixed $F$. So they are marked with the letters showing the type of the states with
index indicating the value of $F$, and the number in a series in front of it all. For example, the
hole state with the lowest energy is of the mixed type---$1hm_1$.

The calculated lower energy levels of electrons  and holes confined in Si NC of diameter in the
range 2.8~nm - 3.3~nm are shown in Fig.~\ref{fig:Chap2_Sec2_Fig9}. The  energy range is limited to
the one of optical pumping used in experiments (2.85~eV). Due to the existence of relatively large
energy separations between some neighboring space quantized levels, one phonon emission processes
will be suppressed. This effect might lead to the so-called "phonon bottleneck" slowing down of
carrier relaxation. And although in different nanocrystals alternative rapid cooling mechanism have
been observed - e.g. CdSe \cite{Hendry_2006} - one can expect that energy relaxation of "hot"
carriers will in our case become slower in comparison to bulk silicon. Thus, an Auger excitation of
erbium ions in silicon dioxide is possible, similar to the impact ionization by hot carriers in
bulk silicon, where it plays a significant role in electroluminescence, but just negligibly affects
the excitation of the erbium photoluminescence due to the fast energy relaxation of hot carriers in
the bulk material.

\subsection{Excitation due to intra-band transition}\label{subsubsec:chap2_sec2_subsubsec1}

The probability of Auger excitation of an erbium ion situated in
SiO$_{2}$ at the distance $a$ from the center of a Si NC as the
result of the transition of a ``hot'' confined carrier from the
state $i$ into the state $i'$ is given by the Fermi golden rule:
\begin{equation}
    W_{i'i}
        = \frac{2\pi}{\hbar}
          \frac{1}{N_f}
          \sum_{ff'}
            \left|\langle f',i'|e\Phi|f,i\rangle\right|^2
            J_T(N)
            \delta(E_{i}-E_{i'}-\Delta_{ff'}-N\hbar\omega_{\mathrm{ph}}),
\label{Fermi1}
\end{equation}
where $\Phi$ is the potential created by the $f$-electron of the
Er$^{3+}$ ion; $f,f'$ enumerate the states of $f$-electrons of the
ion; $\Delta_{ff'}$ is the transition energy; and $N_f$ is the
degeneracy degree of the $f$ state. The integration in the matrix
elements of Eq.~(\ref{Fermi1}) is to be produced over the carriers
confinement space and the f-electron coordinate. Due to the energy
conservation law, the confined carrier transition is accompanied
by the emission of $N$ phonons with energy
$\hbar\omega_{\mathrm{ph}}$.

In the Huang-Rhys model (the model of two displaced oscillators with the same frequency), the
phonon factor $J_T(N)$ is given by \cite{Ridley_book}:
\begin{equation}
    J_T(N)
        = \exp\left[-2S\left(N_{T}+\frac{1}{2}\right)\right]
          \exp\left[\frac{N}{2}\frac{\hbar\omega_\mathrm{ph}}{kT}\right]
          I_{N}\left[ 2S\sqrt{N_{T}\left(N_{T}+1\right)} \right],
\label{phonon factor}
\end{equation}
where $S$ is the Huang-Rhys factor which in the one mode
approximation is given by
\begin{equation}
    S
    = \frac{\varepsilon_{\mathrm{opt}}-\varepsilon_{\mathrm{th}}}{\hbar\omega_{\mathrm{ph}}}
\label{Huang-Rhys}
\end{equation}
with $\varepsilon_{\mathrm{opt}}$ and $\varepsilon_{\mathrm{th}}$
corresponding to the optical and thermal ionization energy,
respectively; $N_T$ is the Bose-Einstein factor:
\begin{equation}
    N_T
        = \frac{1}{\exp\left(\frac{\hbar\omega _{\mathrm{ph}}}{kT}\right)-1}~,
\label{N_T}
\end{equation}
and $I_{N}(x)$ is the modified Bessel function of order $N$.

As the energy levels are highly degenerate, one should average over all the initial states with the
energy $E_i$ and sum over all the final states corresponding to the energy $E_i'$ in
Eq.~(\ref{Fermi1}). All these states are actually split due to nonsphericity of NCs and other
factors. This fact is taken into account by assuming the broadening of levels and adding the value
$\delta E$ to the argument of $\delta$-function in Eq.~(\ref{Fermi1}) and averaging over this value
in the energy range $\Delta_E = 60$~meV (we have  assumed that it is about
 the energy of an optical phonon in bulk Si).
In result Eq.~(\ref{Fermi1}) transforms into
\begin{equation}
    W_{i'i}
    = \frac{2\pi}{\hbar}
      \frac{1}{\Delta_E}
      \frac{1}{N_{i}}
      \sum_{M,M'}
        \frac{1}{N_f}
      \sum_{ff'}
        \left|\langle f'; i',M'|e\Phi|f; i,M \rangle\right|^{2}
        J_T(N),
\label{Fermi2}
\end{equation}
where $N_{i}$ is the degeneracy degree of the initial state $i$.
$M$ and $M'$ enumerate the degenerate states of levels $i$ and
$i'$, and final and initial energies are related through
\begin{equation}
    E_i' \simeq E_i-\Delta_{ff'}-N\hbar\omega_\mathrm{ph}. \label{energy}
\end{equation}

Following the Appendix, one finally gets for the probability of
excitation:
\begin{equation}\label{eq:chap2_sec2_eq3}
W_{i'i}= \frac{\pi}{2\sqrt{\varepsilon_2}} \frac{1}{\tau_{\mathrm{rad}}} \frac{1}{R^4}
\left(\frac{\hbar c}{\Delta_{ff'}}\right)^3 \frac{e^2}{\varepsilon_2^2\Delta_E} I_{i'i}(a) J_T(N),
\end{equation}
where $\varepsilon_{2}$ is the high-frequency dielectric constant in SiO$_{2}$ (as the energy
transmitted processes is much larger than the phonon energies \cite{Abakumov_book_1991}),
$\tau_{\mathrm{rad}}$ the radiative lifetime of the erbium ion in the first excited state
($^{4}I_{13/2}$), $R$ is the radius of the NC, and $I_{i'i}(a)$ are dimensionless factors defined
in the Appendix. From Eq.~(\ref{eq:chap2_sec2_eq3}) one gets using values
$\tau_{\mathrm{rad}}=2$~ms and $R=1.55$~nm:
\begin{equation}\label{eq:chap2_sec2_eq4}
    W_{i'i}
    = 8.3\times10^{9}
    \left(\frac{1.55 \mathrm{nm}}{R}\right)^4
    I_{i'i}(a)
    J_T(N)
    \
    \mathrm{s}^{-1}.
\end{equation}

 It has been shown that $I_{i'i}(a)$ are actually functions of the
relation $a/R$ for holes. In the case of electrons, there is only a weak additional dependence on
the NC size, which one can neglect at least in the NC size range under consideration
\cite{Prokofiev_EMRS_2008}. The results of calculations of the factors $I_{i'i}(a)$ for electrons
and holes are presented in Fig.~\ref{fig:Chap2_Sec2_Fig10}.

 The energy required to get the Er$^{3+}$ ion into the first excited
state $^4\mathrm{I}_{15/2}$ (0.81~meV) with reasonable excess or shortage of energy can be covered
by emission or absorption of phonons. The parameters of the multiphonon transition accompanying the
Auger processes are not well defined. There is no data on existence of the electron-phonon
interaction for Er$^{3+}$ ions in the state $^4$I$_{13/2}$ in SiO$_2$. Thus, the interaction of
confined carriers with optical phonons should be considered. The dispersion of optical phonons in
bulk silicon can be neglected and the multimode model of the phonon transition becomes equivalent
to the one-mode Huang-Rhys model \cite{Abakumov_book_1991}. Phonon factor $J_T(N)$ calculated for
phonon energy equal to 60 meV (which is about the optical phonon energy in Si) is presented in
Table~\ref{tab:chap2_sec2_tab2} calculated with a reasonable value of the Huang-Rhys parameter
$S=0.1$ at room temperature. The exact value of the phonon energy and Huang-Rhys factor S are not
known for the material considered here. We have used value S= 0.1, which is in accordance with
experimental values obtained for Er ions fluorozirconate glasses \cite{Shinn_Sibley_PRB1983}. The
influence of the value of Huang-Rhys parameter is shortly discussed in
Ref.~\cite{Prokofiev_JOL2006}. The interaction with optical phonons is forbidden for electrons in
silicon. One can suppose that the interaction of confined carriers with oxygen vibration could be
also responsible for multiphonon assisted Auger processes in the system under consideration. The
values of phonon factor $J_T(N)$ at $\hbar\omega_\mathrm{ph}=140$~meV corresponding to the oxygen
vibrations are shown in Table~\ref{tab:chap2_sec2_tab2} as well.

In Fig.~\ref{fig:Chap2_Sec2_Fig9} the most effective excitation
processes for NCs with diameter 3.1 nm are demonstrated. The
calculated values of the probabilities for these processes as a
function of the distance between a Er$^{3+}$ ion and the nearest
NC are presented in Fig.~\ref{fig:Chap2_Sec2_Fig11}.

\begin{table}[!h]
\begin{center}
        \begin{tabular}{c||c|c|c|c|c|c|c|c}
            $\boldsymbol{N}$           &  $\boldsymbol{-2}$ &  $\boldsymbol{-1}$ &  $\boldsymbol{0}$  &  $\boldsymbol{+1}$ & $\boldsymbol{+2}$&
            $\boldsymbol{+3}$ &  $\boldsymbol{+4}$ &  $\boldsymbol{+5}$
            \\
            \hline
            \hline
            $\boldsymbol{\hbar\omega_\mathrm{ph}=\phantom{1}60}$~{\bf meV}    & $5.2\times10^{-5}$ &  $0.0096$  &  0.87  &  0.098  & $0.0054$ &
            $0.2\times10^{-3}$ & $5.5\times10^{-6}$ & $1.2\times10^{-7}$
            \\
            \hline
            $\boldsymbol{\hbar\omega_\mathrm{ph}=140}$~{\bf meV}   & $9.0\times10^{-8}$ &  0.0004  &  0.9  &  0.091  &  0.0046 &
            $0.15\times10^{-6}$ & $3.8\times10^{-3}$ & $7.7\times10^{-8}$
            \\
            \hline
        \end{tabular}
\end{center}
\caption{\it \small Phonon factor $J_T(N)$ calculated with $S=0.1$ for two different phonon energies $\hbar\omega_\mathrm{ph}$ and temperature $T=300$~K.}\label{tab:chap2_sec2_tab2}
\end{table}

\subsection{De-excitation of erbium by carriers confined in NCs}\label{subsubsec:chap2_sec2_subsubsec2}

When considering erbium de-excitation due to intra-band transitions of confined carriers, one can
use the same formulae from Sec.~\ref{subsubsec:chap2_sec2_subsubsec1} (and Appendix), but
interchanging  initial and final states, as well as adjusting the phonon factor $J_T(N)$. If erbium
excitation process takes place with carrier transition energy larger than $\Delta_{ff'}=0.81$~eV,
and requires $N$ phonons to be emitted, then the reverse process will be described by $J_T(-N)$
factor, which is much less than $J_T(N)$. That is why most of the processes appear either in
excitation or in de-excitation section.

The difference is not in the factor only.  One should also take
into account that the transitions under consideration take place
between degenerate states. So the total probability is achieved by
summation over the final states and averaging over the initial
ones. As the degeneracy degree can be different for initial and
final states, the probability of excitation and de-excitation
processes can be different even for those of them which do not
need phonons. This difference might be considerable for confined
carriers, especially for holes. Since most of upper levels are
described by larger values of moment $F$, having greater
degeneracy degree, the probability of erbium de-excitation should
be higher for the processes which do not require any phonons to be
emitted or absorbed.

 The most effective de-excitation processes for NCs with diameter
3.1 nm are demonstrated in Fig.~\ref{fig:Chap2_Sec2_Fig9} and calculated values are shown in
Fig.~\ref{fig:Chap2_Sec2_Fig11}.

\subsection{Erbium excitation by the recombination of confined carriers}\label{subsubsec:chap2_sec2_subsubsec3}

Let us consider the excitation of Er$^{3+}$ ions by recombination of confined electron and hole.
For the NCs under consideration ($d\sim 3.1$~nm) the recombination energy is larger than 1.5~eV.
Therefore the energy transfer to the Er$^{3+}$ ion by an Auger recombination of such an exciton can
be effective only if it causes the direct transition of the ion into the third excited state
$^4$I$_{9/2}$ (energy of transition from the ground state $^4$I$_{15/2}$ is $\Delta_{03}=1.55$~eV),
the fourth $^4$F$_{9/2}$ (transition energy $\Delta_{04}=1.9$~eV) or higher excited states. One
should notice that the Er$^{3+}$ ion can practically not be excited directly into the state
$^4$I$_{13/2}$ via such a process, as it should be accompanied by multi-phonon emission with the
number of phonons N $\geq$ 5. The phonon factor $J_T(N)$ is in this case very small (see Table
\ref{tab:chap2_sec2_tab2}). In order to calculate the transition probability we can use
formula~\eqref{Fermi1}, just assuming that initial and final states of confined carriers now belong
to different bands and dividing the probability by the degeneracy of the final state since only one
final state is empty if there is one electron-hole pair in the NC. One should also choose
appropriate parameters of the phonon system.

Crucial for the matrix element evaluation is keeping in mind that the value $\hbar\Delta k$ of
momentum transmitted during recombination process is large. The minima of the conduction band in
$k$-space are shifted from the $\Gamma$~point by the wave vector $k_0=0.85 k_X$ ($k_X$ is the
Brillouin zone edge). And it was shown in Ref.~\cite{Prokofiev_PRB2005} that the momentum to
transmit is even larger than $\hbar k_0$: $\hbar\Delta k=1.15 \hbar k_X$. Such a great momentum can
only be transferred to the $f$-shell of the erbium ion by the Coulomb interaction at a distance
less than the lattice constant of silicon. So the interaction has a contact character and is
determined by the electron and hole wave function values at the position $\mathbf{a}$ of the
erbium.

Once carriers are strongly confined in the NC, and the tunnelling is weak, the interaction is
possible either inside the NC or in its vicinity. When dealing with the Coulomb interaction at
distances smaller than the lattice constant, no screening should be taken into account any more,
and the effective dielectric constant value can be assumed to be $\varepsilon_{\mathrm{eff}}=1$
\cite{Abakumov_book_1991, Prokofiev_PRB2005}. In this case the absolute value square of the matrix
element in Eq.~\eqref{Fermi1} averaged over the degenerate electron and hole states can be
calculated in analogy to bulk Auger processes \cite{Yassievich_SST1993, Moskalenko_PRB2004} as
\begin{equation}\label{Eq:matrix_element_Auger_result1}
    \overline{\left|\langle f',i'|e\Phi|f,i\rangle\right|^2}=
    \frac{(2\pi)^2e^4}{\varepsilon_\mathrm{eff}^2}
     |\langle f|z^2|f'\rangle|^2
     \left|\langle u_0|u_{cz}\rangle\right|^2\left|\xi^{e,i}(\mathbf{a})\right|^2
    \frac{1}{N_{i'}}\sum_{M'}\left|\xi^{h,i'}_{M'0}(\mathbf{a})\right|^2,
\end{equation}
where $\xi^{e,i}(\mathbf{r})$ is the electron envelope function in the initial state and for
shortness of notations the total hole wave function in the final state is written as
$\psi_{FM}^{h,i'}(\mathbf{r})=\sum_m \xi^{h,i'}_{Mm}(\mathbf{r}) u_m$ with $u_m$ $(m=-1,0,+1)$
being the hole Bloch functions; $N_{i'}=2F'+1$ is the degeneracy of the hole state; $\left|\langle
u_0|u_{cz}\rangle\right| \approx 0.25\;$ is the overlap integral between the bottom of the valence
band $\Gamma_{25'}^{l}$ and the second conduction band $\Delta_{2'}^c$ with $k$ at the position in
the first Brillouin zone where the first conduction band has its minimum \cite{Cardona_PR1966}. We
remark, that the quadrupole term plays here the main role in Coulomb interaction. Using expression
\eqref{Eq:matrix_element_Auger_result1} for the right hand side of Eq.~\eqref{Fermi2} divided by
$N_{i'}$ we get the expression for the transfer probability for a given position of Er$^{3+}$ and
radius of the NC
\begin{equation}\label{Eq:Wtrinside}
    W_{\rm tr}(\mathbf{a};R)=\frac{3\pi}{2}\frac{1}{\hbar\Delta_E}
    \left(\frac{e^2}{\varepsilon_\mathrm{eff} R}\right)^2 Q(\mathbf{a};R)
    \left|\langle u_0|u_{cz}\rangle\right|^2 \frac{\gamma_{f}r_f^4}{R^4}
    \;J_T(N).
\end{equation}
Here the factor $\gamma_f r_f^4$ comes from the summation over $f$
and averaging over $f'$ of the absolute value square of the matrix
element $\langle f|z^2|f'\rangle$, where $r_f\approx
0.43$~$\mathring{\mbox{A}}$ is the radius of the $4f$-shell of the
Er$^{3+}$ ion and the unknown factor $\gamma_f$ is of the order
of~1. We have introduced a dimensionless factor $Q(\mathbf{a};R)$
defined by
\begin{equation}\label{Eq:Q_R}
    Q(\mathbf{a};R)=\left(\frac{4\pi}{3}R^3\right)^2\left|\xi^{e,i}(\mathbf{a})\right|^2
          \frac{3}{N_{i'}}\sum_{M'}\left|\xi^{h,i'}_{M'0}(\mathbf{a})\right|^2.
\end{equation}
If we assume the homogeneous probability distribution for the Er$^{3+}$ ion  inside the NC then the
probability of excitation transfer averaged over the position of the Er$^{3+}$ ion  inside the NC
is given by Eq.~\eqref{Eq:Wtrinside} where in place of $Q(\mathbf{a};R)$ we have
\begin{equation}\label{Eq:Q_in}
    Q_\mathrm{in}(R)=\frac{1}{\frac{4\pi}{3}R^3}\int_{a<R}\!\!\mathrm{d}^3\mathbf{a}\;
    Q(\mathbf{a};R).
\end{equation}
In order to calculate the average transfer probability at some distance $D$ from the NC boundary we
introduce
\begin{equation}\label{Eq:Q_surf}
    Q_\mathrm{surf}(R)=\frac{1}{4\pi}\int\!\!\mathrm{d}\Omega\;
    Q(\mathbf{R};R),
\end{equation}
where the integral is taken over the full solid angle $\Omega$.
Then the above-mentioned probability is given by
Eq.~\eqref{Eq:Wtrinside} where in place of $Q(\mathbf{a};R)$ we
have
\begin{equation}\label{Eq:Q_surf_d}
    Q_\mathrm{surf}(R)\exp[-2(\tilde{\kappa}_e+\tilde{\kappa}_h) D/R].
\end{equation}
Here the dimensionless factors
\begin{equation}\label{Eq:Q_surf_d_kappa}
   \tilde{\kappa}_{e(h)}=\sqrt{\frac{2m_{e(h)}^\mathrm{o}[U_{e(h)}-E_{e(h)}] R^2}{\hbar^2}}
\end{equation}
determine the decay of the electron and hole wave functions
outside the NC, $E_{e}$ and $E_{h}$ are the electron and hole
quantization energies, $m_e^\mathrm{o}$ and $m_h^\mathrm{o}$ are
the electron and hole masses outside the NC and $U_e=3.2$~eV and
$U_h=4.3$~eV are the corresponding energy barriers at the NC
boundary. For the considered NCs
$2(\tilde{\kappa}_e+\tilde{\kappa}_h)$ is on the order of $10^2$
(see Table~\ref{Table:interband}). Therefore the probability of
the excitation transfer by the electron-hole recombination decays
rapidly with increase of the distance between the erbium and NC.
It becomes negligible at the distance of only several angstroms.


For local vibrations of erbium ions in fluorozirconate glass by optical transitions from the higher
excited states into the ground state the values around $\hbar\omega_{\mathrm{ph}}=60$~meV and
$S=0.1$ were reported in Ref.~\cite{Shinn_Sibley_PRB1983}. We note that bulk optical phonons in Si
also have approximately the same energy. These values we have used for the calculation of the
phonon factor $J_T(N)$ in the transfer probability (see Table~\ref{tab:chap2_sec2_tab2}). For the
estimation of the transfer probability we have also used $\gamma_f=1$. Then
Eq.~\eqref{Eq:Wtrinside} can be written as
\begin{equation}\label{Eq:W_tr_inside_locale_est}
    W_{\rm tr}=0.8\!\times\! 10^{11}\: Q J_T(N)\; \mathrm{s}^{-1}.
\end{equation}
We have analyzed transitions induced by electron and holes being
in one of the two lowest states. The numerical factors
$Q_\mathrm{in}$ and $Q_\mathrm{surf}$ are given in
Table~\ref{Table:interband} for $R=1.55$~nm together with the
corresponding energies which should be compensated by phonons and
the decay factor $2(\tilde{\kappa}_e+\tilde{\kappa}_h)$.


\begin{table}[!h]
\begin{center}
        \begin{tabular}{c||c|c|c|c|c}
            \bf{Transition} & $\boldsymbol{E_{i'i}-\Delta_{03}}$ & $\boldsymbol{E_{i'i}-\Delta_{04}}$ &  $\boldsymbol{Q_\mathrm{in}}$ & $\boldsymbol{Q_\mathrm{surf}}$ & $\boldsymbol{2(\tilde{\kappa}_e+\tilde{\kappa}_h)}$
            \\
            \hline
            \hline
            $\boldsymbol{1e_0 \rightarrow 1hm_1}$     &  -41~meV  & -391~meV & 1.34 &    0.043 & 98.9
            \\
            \hline
            $\boldsymbol{2e_0 \rightarrow 1hm_1}$     &  186~meV  & -164~meV & 1.24 &    0.098  & 98.0
            \\
            \hline
            $\boldsymbol{1e_0 \rightarrow 1hh_1}$     &  128~meV  & -222~meV & 1.06 &    0.19  & 97.7
            \\
            \hline
            $\boldsymbol{2e_0 \rightarrow 1hh_1}$     &  327~meV  &  -23~meV & 0.77 &    0.20  & 96.8
            \\
            \hline
        \end{tabular}
\end{center}
\caption{\it \small Calculated parameters of several interband
transitions for NC diameter of 3.1~nm.} \label{Table:interband}
\end{table}

From the data presented in Table~\ref{Table:interband} and using Table~\ref{tab:chap2_sec2_tab2},
Eq.~(\ref{Eq:Q_surf_d}) and Eq.~(\ref{Eq:W_tr_inside_locale_est}), one can see that the
electron-hole recombination can effectively excite an Er$^{3+}$ ion situated inside the NC or at a
very short distance from its boundary on a nanosecond or even shorter time scale. However, this
ultrafast excitation process does not lead to an immediate excitation of the first $^4$I$_{13/2}$
excited state of the Er$^{3+}$ ion relevant for the 1.55~$\mu$m emission. Transition to this state
can occur only via a subsequent multiphonon relaxation process which takes place on a microsecond
time scale.

\subsection{Dipole-dipole contribution for the inter-band transitions}\label{subsubsec:chap2_sec2_subsubsec4}

When an erbium ion is situated at a distance $a\gg R$ one can expand the Coulomb interaction
between the recombining electron-hole pair in the NC and the erbium ion in SiO$_{2}$ into the
series over the coordinate of the confined carrier. In the leading order such an approximation
results into the dipole-dipole interaction being the reason for the so-called F\"{o}rster mechanism
of excitation \cite{Foerster}. In this case the large excess momentum of the confined pair is
transferred to the boundary of the NC in the process of recombination, or the recombination is
accompanied by emission of a phonon as in the radiative exciton recombination
\cite{Moskalenko_PRB2007}.

The probability of the excitation governed by the dipole-dipole
interaction can be presented as
\begin{equation}
    W_\mathrm{dd}=\frac{8\pi}{3}\frac{1}{\hbar^2\omega_\mathrm{ph}}
    \frac{e^4}{\varepsilon_{\rm eff}^2 a^6}
    d_{\rm ex}^2
    \sum_{j\geq 3} d_{0j}^2 J_T(N_j),
    \label{Eq:prob_excit_tran_dip_dip1}
\end{equation}
where $J_T(N_j)$ is the phonon factor and $d_{\rm ex}$ is the
dipole momentum of the confined exciton, which can be  estimated
by using its relation to the confined exciton radiative lifetime
\begin{equation}
    \frac{1}{\tau_\mathrm{ex}^\mathrm{rad}}=
        \frac{4e^{2}E_\mathrm{ex}^3d_{\rm ex}^2n_{\mathrm{eff}} }{3\hbar
        ^{4}c^{3}}.
\label{Eqtaurad}
\end{equation}
Here $E_\mathrm{ex}$ is the exciton energy
and the effective refraction index $n_{\mathrm{eff}}$ is determined by the formula from
Ref.~\cite{Thraendhardt_PRB2002}
\begin{equation}
    n_{\mathrm{eff}}=\left(
            \frac{\varepsilon_{\mathrm{m}}}{\varepsilon_\mathrm{eff}}
        \right)^{2}
        {\varepsilon_{\mathrm{m}}}^{1/2},
 \label{refraction index}
\end{equation}
where $\varepsilon_{\mathrm{Si}}$ and $\varepsilon_{\mathrm{m}}$
are the dielectric constants of silicon and medium, respectively,
and $\varepsilon_\mathrm{eff}=(\varepsilon_{\mathrm{Si}} +
2\varepsilon_{\mathrm{m}})/3$.

Matrix elements $d_{0j}$ in \eqref{Eq:prob_excit_tran_dip_dip1} correspond to the transitions in
the $f$-shell of Er$^{+3}$, and can be expressed via the corresponding oscillator strengths
$P_{0j}$:
\begin{equation}
\label{Eq:d_ff'_square_average}
    d_{0j}^2
        = \frac{3\hbar^2}
               {2m_0\Delta_{0j}n_\mathrm{m}}
          P_{0j},
\end{equation}
where, in the simplest approximation, $n_\mathrm{m}$ is the refraction index of the medium
\cite{Judd_PR1962}. To our knowledge, there is no data in the literature concerning the oscillator
strengths of transitions between the levels of the Er$^{3+}$ ion in the considered inhomogeneous
media. However, we can estimate them via the oscillator strengths found for several glasses and
solutions \cite{Miniscalco_JLT1991, Judd_PR1962}: $P_{03}=1-3 \times 10^{-7}$ for transition
$^4$I$_{9/2}\rightarrow$$^4$I$_{15/2}$, $P_{04}\approx 2\times 10^{-6}$ for transition
$^4$F$_{9/2}\rightarrow$$^4$I$_{15/2}$. Based on these data we estimate $d_{03}^2=1\times
10^{-22}$~cm${^2}$, $d_{04}^2=7 \times 10^{-22}$~cm$^2$.

The calculation leads to the estimation
\begin{equation}
    W_\mathrm{dd}
    \lesssim
    10^{-1}
    \left(
        \frac{R}{a}
    \right)^{6}
    \frac{1}{\tau_\mathrm{ex}^\mathrm{rad}}.
 \label{estimation}
 \end{equation}
Thus, one can see that the F\"{o}rster mechanism does not work effectively for the considered
system, especially at some distance from the NC, because the radiative recombination of confined
carriers is a faster process. Again the excitation of the first excited state of the Er$^{3+}$ is
additionally delayed by the multiphonon relaxation from the higher excited states.

\section{Discussion}\label{subsec:chap2_sec2_subsec5}

In the previous section we have shown that the Si-NC mediated excitation of Er$^{3+}$ ions can
proceed by a variety of physical processes. For all of these, efficiency depends on the distance
between Si-NC and Er. Therefore we have generated a simulation of the Er$^{3+}$ distribution as a
function of the distance from the nearest Si NC---see Fig.~\ref{fig:Chap2_Sec2_Fig12}, based on
sample characteristics (Er and NC concentration, NC size), assuming a random distribution of both,
Si NCs and Er$^{3+}$ ions in the SiO$_{2}$ matrix. As can be seen, the vast majority of the
Er$^{3+}$ content is positioned outside the NCs, with only 6.6\% of the Er$^{3+}$ contained inside
the NCs volume. In addition to this statistical prediction, we note that during crystallization
process \cite{Fujii_APL1997}, a considerable part of the Er$^{3+}$ ions statistically present
inside the NCs will become trapped at the Si NC/SiO$_{2}$ boundary. These Er$^{3+}$ ions are
susceptible of an instantaneous excitation directly into the first excited state, via direct
absorption of photons with enough energy to excite erbium and create an electron-hole pair in the
NC. In silicon the absorption processes are usually accompanied by phonon emission due to indirect
band structure. When an Er$^{3+}$ ion is located inside a NC the absorption can accompanied by an
excitation of this ion instead of phonon emission. This process is not considered in the
theoretical part of this paper. We also note that the Er$^{3+}$ ions remaining in the NC will
induce a donor center, as a result of which they will de-excite non-radiatively very fast
\cite{Prokofiev_PRB2005}.

As we can see from Fig.~\ref{fig:Chap2_Sec2_Fig4}, in panel {\bf b}, the fast excitation of
Er$^{3+}$ ions is completed within 20 ns after the excitation pulse, and we have estimated in our
previous contribution \cite{Izeddin_PRL2006} that $\sim$ 50\% of the Er$^{3+}$ content is involved
in this fast process. The observed fast (sub-microsecond) Er-related PL is only possible when
Er$^{3+}$ is excited directly into the first excited state $^{4}$I$_{13/2}$. From
Fig.~\ref{fig:Chap2_Sec2_Fig12} one can see that approximately 10 \% of Er$^{3+}$ ions occur at a
distance less than 1.15R from the NCs' centers, where according Fig.~\ref{fig:Chap2_Sec2_Fig11} the
most effective carrier cooling processes can provide Er$^{3+}$ excitation at times shorter than 20
ns. However, we should point out that the presented calculations of transition probabilities serve
mostly as demonstration that such a fast mechanism is physically feasible. The theoretical
considerations are valid for a single hole/single electron transition and for strictly spherical Si
quantum dots. In this case, strong selection rules appear and only matrix elements of some higher
multipoles of the electric field potential lead to considerable non-vanishing contributions,
causing a strong decrease of the probability with increasing distance from Er$^{3+}$ to Si NCs.
{\it E.g.}, the Auger process accompanied by the 3{\it e}$_{1}$ $\rightarrow$ 2{\it e}$_{0}$
transition is dominated by the quadrupole-dipole interaction (the dipole is related to Er$^{3+}$).
Therefore one expects that the real probabilities might be higher than the calculated ones, and
will decrease less abruptly with distance, due to non-sphericity of Si NCs. Even more important is
the possibility of multiple exciton generation in a single NC, also not accounted for in the
calculations. Under the excitation conditions used in the experiment, we estimate that on average
about 5 electron-hole pairs are created per Si NC. On one side it will directly increase the
excitation probability of Er$^{3+}$ by a given NC at the initial stage. On the other side the
strong exciton-exciton interaction will then generate very hot carriers leading to population of
higher excited states whose participation has not been considered in Section IV. This will result
in much higher transfer rates, due to a higher electron state localization outside Si NC. All these
effects will contribute to the ``fast'' PL and improve the agreement with our experimental results.

Previously, in \cite{Izeddin_PRL2006}, we have estimated that $\sim$ 0.5\% of the total Er$^{3+}$
decay radiatively ($\tau \sim$ 3 ms) after excitation. From these 0.5\% of ions, $\sim$ 50\% have
been excited due to the fast process, but have not been de-excited by the inverse mechanism: the
``residual concentration". The other $\sim$ 50\% ($\sim$ 0.25\% of the total Er$^{3+}$ content in
the sample) have been excited due to inter-band recombination of the carriers in the NC. In our
theoretical considerations we have shown that the excitation via the Auger process, accompanied by
recombination of an exciton in the Si NC, can only take place via the contact. The F\"{o}rster
mechanism is not effective as shown in Sec. IVC. In any case the excitation of Er$^{3+}$ ion
accompanied by a recombination of an electron-hole pair takes place into the second or the third
excited state of the 4f shell of Er$^{3+}$. The characteristic excitation of the 1.5 $\mu$m
Er$^{3+}$ PL is, in this case, a two-step process with time constant $\tau$ = $\tau_{1}$ +
$\tau_{2}$, where $\tau_{1}$ is the characteristic Er excitation into one of the higher excited
states by the band-to-band recombination of confined carriers, on the nanosecond or even shorter
time scale, and $\tau_{2}$ $\approx$ 1.2 $\mu$s is the relaxation time of the excited Er$^{3+}$
from a higher into the first excited state $^{4}$I$_{13/2}$. This process is responsible for the
$\mu$s rise of the Er-related PL signal.

Following the described model, one expects that the {\it fast-to-slow} Er$^{3+}$ PL ratio will
depend on the number of ``hot" carriers confined in the Si NCs: a high number of ``hot" carriers
will favor the fast intra-band excitation process and thus increase the relative importance of the
``fast" Er$^{3+}$ PL. This is indeed confirmed in Fig.~\ref{fig:Chap2_Sec2_Fig5}, where we can see
that under excitation conditions where creation of ``hot" carriers is more likely, \textit{i.e.}
excitation with higher energy photons or high flux pumping and subsequent multiple carrier
generation in the NCs, the fast-to-slow PL intensity ratio of Er PL increases. In the case of
formation of several electron-hole pairs per Si NC, these can undergo a quick Auger recombination
process, and in that way the excess pairs escape from participating in the ``slow" excitation
transfer process.

Therefore, the following sequence of excitation and de-excitation
processes can be proposed:
\begin{enumerate}
\item After the laser pulse, an Auger-like process of fast excitation
takes place, by intra-band relaxation of ``hot'' carriers transferring their excess energy directly
into the first excited state of Er$^{3+}$ ion. Up to $\sim$ 50\% of all the Er$^{3+}$ content is in
the effective range of this interaction, before the de-excitation processes start to decrease the
number of excited ions. We note that this percentage is dependent on the erbium concentration as
indeed observed experimentally in \cite{Imakita_EPJD2005}.
\item Fast de-excitation of Er$^{3+}$ takes place by a reverse process,
transferring energy to carriers confined in the NCs. This process must be phonon assisted when the
involved transitions do not match the energy conservation requirements; which will manifest in a
slower PL decay and temperature dependence of the characteristic decay constant. The ``hot"
carriers in the NCs undergo an intra-band relaxation accompanied by phonon emission, which reduces
the number of carriers available for the ``quick" excitation process. The rate of relaxation
processes should increase with temperature.

A small percentage of the excited Er$^{3+}$ ions will overcome the
fast de-excitation, giving rise to the ``residual concentration"
of Er$^{3+}$ which is excited in the nanoseconds time window, and
de-excites radiatively. This is revealed in the Er$^{3+}$ PL
kinetics as the non-zero origin of the microsecond rise and
subsequent radiative decay of Er$^{3+}$ PL.
\item Once the confined carriers have cooled down to the bottom
(top) of the conduction (valence) band, there is a probability of
Er$^{3+}$ excitation by recombination, to the upper Er excited
states, as described above. This probability exists also on
shorter time scales, but the number of Er$^{3+}$ ions which can be
accessed is low, and therefore this inter-band process is covered
by the intra-band excitation process.
\end{enumerate}

Based on the presented experimental data and theoretical
modelling, we conclude that three types of optically active
Er$^{3+}$ ions coexist in the investigated material:
\begin{itemize}
\item Type 1: Er$^{3+}$ ions that can only be excited resonantly,
under direct excitation. These ions have predominantly radiative
decay---like Er$^{3+}$ in SiO$_{2}$---and are at large enough
distance from the Si NCs to prevent their interaction.

\item Type 2: Er$^{3+}$ ions that are excited by energy transfer
from inter-band recombination of excitons in Si NCs, at the microseconds time regime and which
decay predominantly radiatively ($\tau$ $\sim$ 2-3 ms), independent of temperature. This type of
Er$^{3+}$ ions are those accounted for in the usual estimations of optical activity measurements
and constitute about 0.25\% to 0.5\% of the total Er$^{3+}$ content.

\item Type 3: Er$^{3+}$ ions excited via the intra-band transition
of carriers in the NCs and with very strong non-radiative quenching, responsible for the sub-$\mu$s
PL described above, whose properties mirror those of Er$^{3+}$ in crystalline-Si ($\sim$ 50\% of
Er$^{3+}$). A small percentage of these ions, will overcome the fast non-radiative de-excitation,
and will be also accounted for in optical activity measurements.
\end{itemize}

Finally, for the sake of completeness we note that in addition to
these, there could also be Er dopants which are not optically
active due to, \textit{e.g.} precipitation.

An independent confirmation of the proposed excitation model of Er$^{3+}$ ions by intra-band
transitions of confined carriers in the Si NC is indeed given by Fig.~\ref{fig:Chap2_Sec2_Fig8}.
There, for excitation energy higher than E$_{th}$ $\approx$ 2.6 - 2.7 eV, a second excitation
mechanism (due to intraband carrier cooling) is enabled, increasing the ratio between absorption
and effective excitation cross section of Er PL. This threshold energy is sufficient to create a
``hot" carrier that can excite Er$^{3+}$ directly into the first excited state by cooling into the
bottom of the conduction band (or the top of the valence band) \cite{NP}.

\section{Conclusions}\label{subsec:chap2_sec2_subsec6}
With the results of this study, an important puzzle concerning the mechanism of excitation of
Er$^{3+}$ by Si NCs has been solved. We have shown that the ``missing" dopants that were apparently
losing optical activity upon doping with Si NCs, and which did not contribute to PL, are actually
very efficiently excited by the Si NCs, via an intra-band Auger transfer process, but undergo also
a very effective excitation back-transfer process. The back-transferred carriers can again excite
the Er$^{3+}$ ion or escape from being available for the intra-band excitation process due to a
thermalization and Auger recombination processes inside a NC. In particular, we point out that the
results of this study clearly show that in the first microsecond after the excitation laser pulse,
a vast majority of Er dopants attain the excited state; that implies that in that short time window
the population inversion is reached, in that way fulfilling a necessary although in itself
insufficient condition for realization of optical gain and laser. Future research will tell,
whether the SiO$_2$:(Er, Si NCs) material can be engineered in such a way that the sensitization of
Er emission is realized without the detrimental effect of reduction of the concentration of
Er$^{3+}$ ions, with the temperature-stable, predominantly radiative recombination. This study
shows that achieving that will require careful and simultaneous optimization of Er$^{3+}$ and NC
concentrations, NC size and size distribution, and, very importantly, very precise tuning of Er -
NC distance, on a nanometer scale.

On the other hand, we point out that fast recombination kinetics
reported here leads to the effective recombination rate of
Er$^{3+}$ in the environment of Si NCs exceeding by 2 orders of
magnitude the fastest quenching rate (due to the Auger process
involving free carriers) reported, to our knowledge, for Er in
crystalline Si. This very efficient PL quenching of Er-related PL
might be explored for GHz modulation of the 1.5 $\mu$m emission
from Er-doped structures.

\newpage
\appendix*
\section{Calculations for the excitation due to intra-band transition}\label{app:app1}

When dealing with the Coulomb interaction between an $f$-electron of an Er$^{3+}$ ion situated in
SiO$_2$ and a carrier confined in Si NC, one should take into account the difference in dielectric
constant values of Si ($\varepsilon_1=12$) and SiO$_2$ ($\varepsilon_2=2$). We note that the Auger
process is determined by the high frequency dielectric constant, as the transition energy
$\Delta_{ff'}$ is much larger than the lattice vibration energy~\cite{Abakumov_book_1991}.

The potential created by a point charge $q$ at the distance
$\mathbf{a}$ from the center of the sphere of radius $R$ ($R<a$)
with dielectric constant $\varepsilon_1$ in the media with
dielectric constant $\varepsilon_{2}$ is obtained as a solution to
the Poisson equation and is given by the equations:
\begin{equation}
\Phi_{1}(\mathbf{r},\mathbf{\mathbf{a}})=\frac{q}{\varepsilon_{2}a}\left[1+\sum_{l=1}^{\infty}
\left(\frac{r}{a}\right)^l\frac{(2l+1)\varepsilon_{2}}{l\varepsilon_{1}+(l+1)\varepsilon_{2}}P_l(\cos\vartheta)\right],
\label{potential1}
\end{equation}
inside the sphere ($r<R$), and
\begin{equation}
\Phi_{2}(\mathbf{r},\mathbf{\mathbf{a}})=\frac{q}{\varepsilon_{2}|\mathbf{r
-\mathbf{a}}|}-\frac{q(\varepsilon_{1}-\varepsilon_{2})}{\varepsilon_{2}r}
    \sum_{l=1}^{\infty}
\left(\frac{R}{a}\right)^{l+1}\frac{l}{l\varepsilon_{1}+(l+1)\varepsilon_{2}}\left(\frac{R}{r}\right)^{l}P_l(\cos\vartheta),
\label{potential2}
\end{equation}
outside the sphere ($r>R$), where $\vartheta$ is the angle between $\mathbf{r}$ and $\mathbf{\mathbf{a}}$:
$\cos(\vartheta)=({\mathbf{r,\mathbf{\mathbf{a}}})/ra}$. Note that the Coulomb potential in the case of interaction of two charges is given by $\frac{1}{2} \left[ \Phi(\mathbf{r}, \mathbf{a}) + \Phi(\mathbf{a}, \mathbf{r}) \right]$, but it is easy to show that in our case
$\Phi(\mathbf{r}, \mathbf{a}) = \Phi(\mathbf{a}, \mathbf{r})$.

Let us consider the potential $\Phi$ in Eqs.~(\ref{Fermi1}),
(\ref{Fermi2}). Introducing the coordinate $\mathbf{r}'$ related
to the center of the ion ($r'\lesssim r_f$, where $r_f$ is the
size of the $f$-shell) and using the fact that $r_f \ll a$, the
potential can be expanded into a series by $\mathbf{r}'$ taking
into account the linear term only:
\begin{equation}
    \Phi(\mathbf{r},\mathbf{\mathbf{a}}+\mathbf{r}')
    \approx
    \Phi(\mathbf{r},\mathbf{\mathbf{a}})
    +
    \frac{\partial \Phi(\mathbf{r},\mathbf{\mathbf{a}})}
        {\partial \mathbf{a}}
    \mathbf{r}'.
\label{Phi_exp}
\end{equation}
In order to use formulae \eqref{potential1} and \eqref{potential2} the integration of matrix
elements in Eq.~(\ref{Fermi2}) should be produced over $\mathbf{r}'$ for $0<r'<r_f$. Relatively
high energy barriers at the boundary of a NC (3.2 and 4.3~eV for electrons and holes, respectively)
allow only a small portion of the confined carriers charge density to penetrate outside.  The
charge density of confined carriers occurring in SiO$_{2}$ due to tunnelling is just a few
percents~\cite{Moskalenko_PRB2007}. Therefore the largest contribution is given by the Coulomb
interaction induced by the carrier density inside the NC. To that end, it is enough to use
potential $\Phi_1$ (see Eq.~\ref{potential1}) only. We can write for it

\begin{equation}
    \frac{\partial \Phi_1(\mathbf{r},\mathbf{\mathbf{a}})}
         {\partial \mathbf{a}}
    =
    \frac{q}{a^2 \varepsilon_2} \mathbf{J},
\label{Phi_1_derivative}
\end{equation}
where
\begin{equation}
    \mathbf{J}=-\frac{\mathbf{\mathbf{a}}}{a}J_{1}+\frac{\mathbf{r}}{r}J_{2},
\label{factorJ}
\end{equation}
and
\begin{equation}
J_{1} = 1 + \sum_{l=1}^{\infty}
    \frac{(2l+1)\varepsilon_{2}}{l\varepsilon_{1}+(l+1)\varepsilon_{2}}
    \left(\frac{r}{a}\right)^l
    \left[(l+1)P_{l}(\cos\vartheta)+\cos\vartheta \frac{\partial P_{l}(\cos\vartheta)}{\partial \cos\vartheta}\right],
\label{factorJ1}
\end{equation}
\begin{equation}
J_{2} =
    \sum_{l=1}^{\infty}
    \frac{(2l+1)\varepsilon_{2}}{l\varepsilon_{1}+(l+1)\varepsilon_{2}}
    \left(\frac{r}{a}\right)^{l}
    \frac{\partial P_{l}(\cos\vartheta)}{\partial \cos\vartheta}.
\label{factorJ2}
\end{equation}
Then Eq.~(\ref{Fermi2}) transforms into
\begin{equation}
W_{i'i}
    =
    \frac{2\pi e^4}{\hbar\Delta_E\varepsilon_2^2 a^4}
    \frac{1}{N_{i}}
    \sum_{M,M'}
        \frac{1}{N_f}
        \sum_{ff'}
            \left|\mathbf{d}_{ff'} \langle i',M'|\mathbf{J}|i,M \rangle\right|^{2} J_T(N),
\label{Fermi3}
\end{equation}
where the ion dipole momentum is given by
\begin{equation}
    \mathbf{d}_{ff'}
    = \int
        \psi_{f'}^*(\mathbf{r})
        \mathbf{r}
        \psi_f(\mathbf{r})
    \mathrm{d}^3 \mathbf{r}.
\label{dipole}
\end{equation}
Averaging Eq.~(\ref{Fermi3}) over the directions of
$\mathbf{d}_{ff'}$ one gets
\begin{equation}
W_{i'i}  =  \frac{2\pi e^4}
        {3\varepsilon_2^2\hbar\Delta_E R^4}
    \frac{1}{N_f}
    \sum_{ff'}
        {|\mathbf{d}_{ff'}|}^{2}
    I_{i'i}(a)
    J_T(N),
\label{Fermi4}
\end{equation}
where the dimensionless factor $I_{i'i}(a)$ is defined by
\begin{equation}
    I_{i'i}(a)
    =   \frac{1}{N_i}
        \left(\frac{R}{a}\right)^{4}
        \sum_{M,M'}
            |\langle i',M' | \mathbf{J} | i,M \rangle|^2,
\label{I_ii}
\end{equation}
where the square of the matrix element absolute value is assumed
to be averaged over the directions of vector $a$. Introducing the
radiative lifetime $\tau_{\mathrm{rad}}$ of the erbium ion in the
first excited state
($^4$I$_{13/2}$)~\cite{Suchocki_Langer_PRB1989}
\begin{equation}
    \frac{1}{\tau_{\mathrm{rad}}}=\frac{1}{N_{f'}}
    \sum_{ff'}\frac{4}{3}
    e^{2}d_{ff'}^2
        \frac{\sqrt{\varepsilon_2}(\Delta_{ff'})^3}{\hbar^4 c^3},
\label{tau_rad_Dffp}
\end{equation}
one gets finally Eq.~(\ref{eq:chap2_sec2_eq3}). The results of calculations of the factors
$I_{i'i}(a)$ for electrons and holes are given in Fig.~\ref{fig:Chap2_Sec2_Fig10}.

For completeness, we consider the contribution of carrier density outside the dot to
the probability of Er$^{3+}$ excitation. The potential $\Phi_{2}$ given by Eq.~(\ref{potential2}) should be
used outside the NC. We get
\begin{equation}
    \frac{\partial \Phi_2(\mathbf{r},\mathbf{\mathbf{a}})}
        {\partial \mathbf{a}}
        =
        \frac{q}{R^2 \varepsilon_2} \mathbf{J'},
\label{Phi_2_derivative}
\end{equation}
where
\begin{equation}
    \mathbf{J'}
        =
        \frac{\mathbf{\mathbf{a}}R}{a^2} J'_{1}
        -\frac{\mathbf{r}R}{ar}J'_{2}
        +\frac{(\mathbf{r}-\mathbf{a})R^2}{|\mathbf{r -\mathbf{a}}|^3},
\label{j_ii}
\end{equation}
\begin{equation}
J'_{1} = \sum_{l=1}^{\infty}
    \frac{\varepsilon_{1}-\varepsilon_{2}}{l\varepsilon_{1}+(l+1)\varepsilon_{2}}
    \left(\frac{R}{a}\right)^{(l+1)}
    \left(\frac{R}{r}\right)^{(l+1)}
    \left[(l+1)P_{l}(\cos\vartheta)+\cos\vartheta \frac{\partial P_{l}(\cos\vartheta)}{\partial \cos\vartheta}\right],
\label{factorJ1prime}
\end{equation}
\begin{equation}
J'_{2} =
    \sum_{l=1}^{\infty}
    \frac{\varepsilon_{1}-\varepsilon_{2}}{l\varepsilon_{1}+(l+1)\varepsilon_{2}}
    \left(\frac{r}{a}\right)^{l}
    \frac{\partial P_{l}(\cos\vartheta)}{\partial \cos\vartheta}\;.
\label{factorJ2prime}
\end{equation}
Producing the calculations analogous to the ones described above
for the carriers being inside the NC, one can find that the
contribution of the confined carriers tunnelling to the excitation
probability is given by an expression similar to
Eq.~(\ref{Fermi4}). Our calculations have shown that the input is
negligible.


\begin{center}
\newpage \bigskip {\Large FIGURE CAPTIONS\vspace{1cm}}
\end{center}

\begin{figure}[!h]
\caption{Time-integrated PL spectra at T = 10 and 300 K, and excitation wavelength $\lambda_{exc}$ = 450 nm. Si NCs, Er-related, and a third (at low T) PL bands can be observed. The Si NC-related band is blown up for clarity. The arrows show PL peaks related to emission from higher excited states of the Er$^{3+}$ ion, superimposed to the NCs excitonic PL.}
\label{fig:Chap2_Sec2_Fig1}
\end{figure}
\begin{figure}[!h]
\caption{Time-resolved contour plot of the PL spectrum of the Si NCs excitonic-related band, in the microseconds ({\bf a}) and in the sub-microsecond ({\bf b}) time ranges. The PL intensity (color contour plots) is represented as a function of time (X axis) and detection wavelength (Y axis). Panel {\bf a} also shows the spectrum recorded at t = 100 $\mu$s, and the PL decay at $\lambda$ = 850 nm (sections from the contour plot). Drift of the band maximum toward longer wavelength is observed for longer time delay in panel {\bf a}. Note the difference between the center of the bands in both figures ($\approx$ 200 meV).}
\label{fig:Chap2_Sec2_Fig2}
\end{figure}
\begin{figure}[!h]
\caption{Room temperature PL decay kinetics recorded at $\lambda$ = 1200 nm (maximum of the PL broad band shown in Fig. \ref{fig:Chap2_Sec2_Fig1}) and $\lambda$ = 1535 nm. In the inset, the spectrum recorded at t = 200 ns after the laser excitation pulse, at T = 10 K, is shown.}
\label{fig:Chap2_Sec2_Fig3}
\end{figure}
\begin{figure}[!h]
\caption{In panel {\bf a}: Si NCs PL kinetics, recorded at
$\lambda$ = 860 nm, and room temperature; time resolution of the
system was 2 ns. Excitation was provided with a 5 ns pulse, at
$\lambda _{exc}$ = 450 nm. Initial nanoseconds decay is stretched
and followed by a final decay with characteristic lifetime of
thick microseconds. In panel {\bf b}: Er-related 1.5 $\mu$m PL
kinetics for the first microsecond after the excitation pulse is
shown, at T = 10 K and RT.} \label{fig:Chap2_Sec2_Fig4}
\end{figure}
\begin{figure}[!h]
\caption{Er-related PL kinetics, in double logarithmic scale for different excitation conditions, at RT. Intensity has been normalized to the maximum of the `slow' component, in order to compare the fast-to-slow intensity ratio. In panels ({\bf a}) and ({\bf b}) the excitation wavelength (450 nm) has been kept constant, and flux has been decreased about one order of magnitude. In panels ({\bf b}) and ({\bf c}), flux has been kept in the same order of magnitude, and wavelength excitation has been increased to 650 nm.}
\label{fig:Chap2_Sec2_Fig5}
\end{figure}
\begin{figure}[!h]
\caption{Flux dependence of Er$^{3+}$ and Si NCs PL for several excitation wavelengths is shown,
recorded at RT.} \label{fig:Chap2_Sec2_Fig6}
\end{figure}
\begin{figure}[!h]
\caption{PL excitation cross section of Er$^{3+}$ and Si NCs PL as a function of excitation
wavelength, which have been determined by using formula (\ref{eq:chap2_sec2_eq1}) to fit the curves
of Fig.~\ref{fig:Chap2_Sec2_Fig6}.} \label{fig:Chap2_Sec2_Fig7}
\end{figure}
\begin{figure}[!h]
\caption{Er- and Si NC-related effective excitation cross section $\sigma_{PL}$ as a function of
the optical absorption coefficient $\alpha$, for each given excitation wavelength.}
\label{fig:Chap2_Sec2_Fig8}
\end{figure}
\begin{figure}[!h]
\caption{Electrons and holes energy levels in Si NCs in SiO$_{2}$, as a function of NC diameter. On
the right-hand side the Er$^{3+}$ energy levels are presented. The most effective erbium excitation
and de-excitation processes due to intra-band carrier transitions are shown.}
\label{fig:Chap2_Sec2_Fig9}
\end{figure}
\begin{figure}[!h]
\caption{Factors $I_{ii'}$ for erbium excitation/de-excitation by
electron intra-band transitions as a function of a/R, where a is
the distance of the Er$^{3+}$ ion from the center of the NC and R
the radius of the NC.} \label{fig:Chap2_Sec2_Fig10}
\end{figure}
\begin{figure}[!h]
\caption{The probability of the most effective excitation and
de-excitation processes as a function of a/R (a is the erbium
distance from the center of a NC and R the radius of the NC) for a
NC with diameter 3.1 nm.} \label{fig:Chap2_Sec2_Fig11}
\end{figure}
\begin{figure}[!h]
\caption{A simulation of the distance distribution of Er$^{3+}$ ions to the center of their nearest
neighboring NC assuming a random distribution of both: 6.6\% of the Er$^{3+}$ ions are contained
inside NCs.} \label{fig:Chap2_Sec2_Fig12}
\end{figure}


\newpage


\begin{references}

\bibitem{Kenyon_JPM1994} A. J. Kenyon, P. F. Trwoga, M. Federighi, and C. W. Pitt, J. Phys.: Condens.
Matter {\bf 6}, {\bf L}319 (1994)
\bibitem{Fujii_APL1997} M Fujii, M. Yoshida, Y. Kanzawa, S. Hayashi, and K. Yamamoto, Appl. Phys. Lett. {\bf 71}, 1198 (1997)
\bibitem{Chryssou_Kenyon_APL1999} C. E. Chryssou, A. J. Kenyon, T. S. Iwayama, C. W. Pitt, and D. E. Hole, Appl. Phys. Lett. {\bf 75} 2011 (1999)
\bibitem{Pacifici_PRB2003} D. Pacifici, G. Franzò, F. Priolo, F. Iacona, and L. Dal Negro Phys. Rev. B {\bf 67}, 245301 (2003)
\bibitem{Kik_Polman_APL2000} P. G. Kik, M. L. Brongersma, and A. Polman, Appl. Phys. Lett. {\bf 76}, 2325 (2000)
\bibitem{Kovalev_PRL1998} D. Kovalev, H. Heckler, M. Ben-Chorin, G. Polisski, M. Schwartzkopff, and F. Koch, Phys. Rev. Lett. {\bf 81}, 2803 (1995)
\bibitem{Castagna_Coffa_PE2003} M. E. Castagna, S. Coffa, M. Monaco, L. Caristia, A. Messina, R. Mangano, and C. Bongiorno, Physica E {\bf 16}, 547 (2003)
\bibitem{Nazarov_Skorupa_APL2005} A. Nazarov, J. M. Sun, W. Skorupa, R. A. Yankov, I. N. Osiyuk, I. P. Tjagulskii, V. S. Lysenko, and T. Gebel, Appl. Phys. Lett. {\bf 86}, 151914 (2005)
\bibitem{Walters_Atwater_NatMat2005} R. J. Walters, G. I. Bourianoff, and H. A. Atwater, Nature Mat. {\bf 4}, 143 (2005)
\bibitem{Foerster} T. F\"{o}rster, Ann. Physik {\bf 2}, 55 (1948)
\bibitem{Dood_PRB2005} M. J. A. de Dood, J. Knoester, A. Tip, and A. Polman, Phys. Rev. B {\bf 71}, 115102 (2005)
\bibitem{Agranovich_1982} V. M. Agranovich and M. D. Galanin, {\it Electronic Excitation Energy Transfer in Condensed Matter}, Elsevier (1982)
\bibitem{Wu_PRL2001} X. L. Wu, Y. F. Mei, G. G. Siu, K. L. Wong, K. Moulding, M. J. Stokes, C. L. Fu, and X. M. Bao, Phys. Rev. Lett. {\bf 86}, 3000 (2001)
\bibitem{Senter_Coffer_PRL2004} R. A. Senter, C. Pantea, Y. Wang, H. Liu, T. W. Zerda, and J. L. Coffer, Phys. Rev. Lett. {\bf 93}, 175502 (2004)
\bibitem{Fuji_JAP2004} M. Fujii, K. Imakita, K. Watanabe, and S. Hayashi, J. Apl. Phys. {\bf 95},
272 (2004)
\bibitem{Savchyn_PRB2007} O. Savchyn, F. R. Ruhge, P. Kik, R. M. Todi, K. R. Coffey, H. Nukala, and
H. Heinrich, Phys. Rev. B {\bf 76}, 195419 (2007)
\bibitem{Imakita_Fujii_PRB2005} K. Imakita, M. Fujii, and S. Hayashi, Phys. Rev. B {\bf 71}, 193301 (2005)
\bibitem{Lee_LT2005} J. Lee, J. H. Shin, and N. Park, J. Lightwave Technol. {\bf 23}, 19 (2005)
\bibitem{Wojdak_PRB2004} M. Wojdak, M. Klik, M. Forcales, O. B. Gusev, T. Gregorkiewicz, D. Pacifici, G. Franzò, F. Priolo, and F. Iacona, Phys. Rev. B {\bf 69}, 233315 (2004)
\bibitem{Kik_Polman_JAP2000} P. G. Kik and A. Polman, J. Appl. Phys. {\bf 88}, 1992 (2000)
\bibitem{Izeddin_PRL2006} I. Izeddin, A. S. Moskalenko, I. N. Yassievich, M. Fujii, and T. Gregorkiewicz, Phys. Rev. Lett. {\bf 97}, 207401 (2006)
\bibitem{Moskalenko_PRB2007} A. S. Moskalenko, J. Berakdar, A. A. Prokofiev, and I. N. Yassievich, Phys. Rev. B {\bf 76}, 085427 (2007)
\bibitem{Kenyon_JAP2004} A. J. Kenyon, C. E. Chryssou, C. W. Pitt, T. Shimizu-Iwayama, D. E. Hole, N. Sharma, and J. Humphreys, J. Appl. Phys. \textbf{91}, 367 (2002)
\bibitem{Schenk_JAP1997} A. Schenk and G. Heiser, J. Appl. Phys. {\bf 81}, 7900 (1997)
\bibitem{Chelikowsky_PRB1977} J. R. Chelikowsky and M. Schlueter, Phys. Rev. B {\bf 15}, 4020 (1977)
\bibitem{Abakumov_book_1991} V. N. Abakumov, V. I. Perel, and I. N. Yassievich, {\it Nonradiative Recombination in Semiconductors}, Elsevier (1991)
\bibitem{Hendry_2006} E. Hendry, M. Koeberg, F. Wang, H. Zhang, C. de Mello Doneg\'{a}, D.
Vanmaekelbergh, and M. Bonn, Phys. Rev. Lett. {\bf 96}, 057408 (2006)
\bibitem{Ridley_book} B.~K.~Ridley, {\it Quantum Processes in Semiconductors}, (Clarendon Press,
Oxford) (1982)
\bibitem{Prokofiev_EMRS_2008} A. A. Prokofiev, A. S. Moskalenko, and I. N. Yassievich, Mater. Sci. and Eng. B {\bf 146}, 121 (2008)
\bibitem{Shinn_Sibley_PRB1983} M. D. Shinn, W. A. Sibley, M. G. Drexhage, and R. N. Brown, Phys. Rev. B {\bf 27}, 6635 (1983)
\bibitem{Prokofiev_JOL2006} A. A. Prokofiev et al., J. of Lum. {\bf 121}, 222 (2006)
\bibitem{Prokofiev_PRB2005} A. A. Prokofiev, I. N. Yassievich, H. Vrielinck, and T. Gregorkiewicz, Phys. Rev. B {\bf 72}, 045214 (2005)

\bibitem{Yassievich_SST1993} I. N. Yassievich, and L. C. Kimerling, Semicond. Sci. Technol. {\bf 8}, 718 (1993)
\bibitem{Moskalenko_PRB2004} A. S. Moskalenko, I. N. Yassievich, M. Forcales, M.Klik, and T. Gregorkiewicz, Phys. Rev. B {\bf 70}, 155201 (2004)
\bibitem{Cardona_PR1966} M. Cardona and F.H. Pollak, Phys. Rev. {\bf 142}, 530 (1966)

\bibitem{Thraendhardt_PRB2002} A. Thraendhardt, C. Ell, G. Khitrova, and H. M. Gibbs, Phys. Rev. B {\bf 65}, 035327 (2002)
\bibitem{Judd_PR1962} B. R. Judd, Phys. Rev. {\bf 127}, 750 (1962)
\bibitem{Miniscalco_JLT1991} W. J. Miniscalco, J. Lightwave Techn. {\bf 9}, 234 (1991)
\bibitem{Imakita_EPJD2005} K. Imakita, M. Fujii, and S. Hayashi, Eur. Phys. J. D {\bf 34}, 161
(2005)
\bibitem{NP} D. Timmerman, I. Izeddin, P. Stallinga, I.N. Yassievich, and T. Gregorkiewicz, Nature Photonics 2, 105 (2008).


\bibitem{Suchocki_Langer_PRB1989} A. Suchocki, and J. M. Langer, Phys. Rev. B {\bf 39}, 7905 (1989)


\end{references}
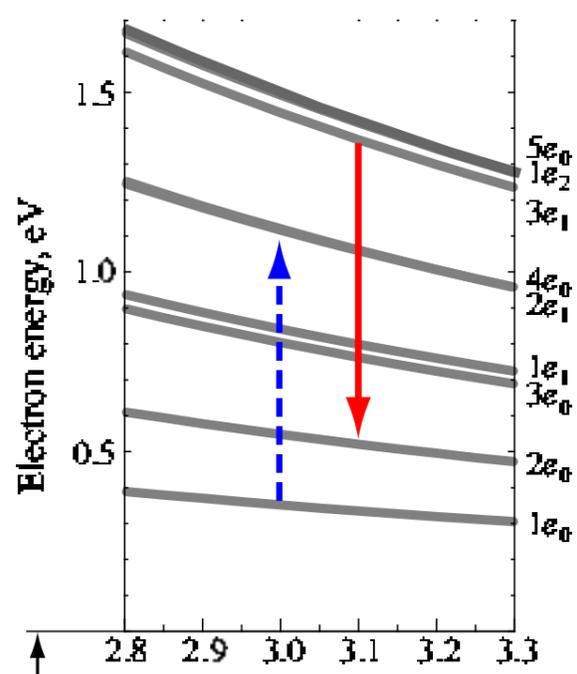
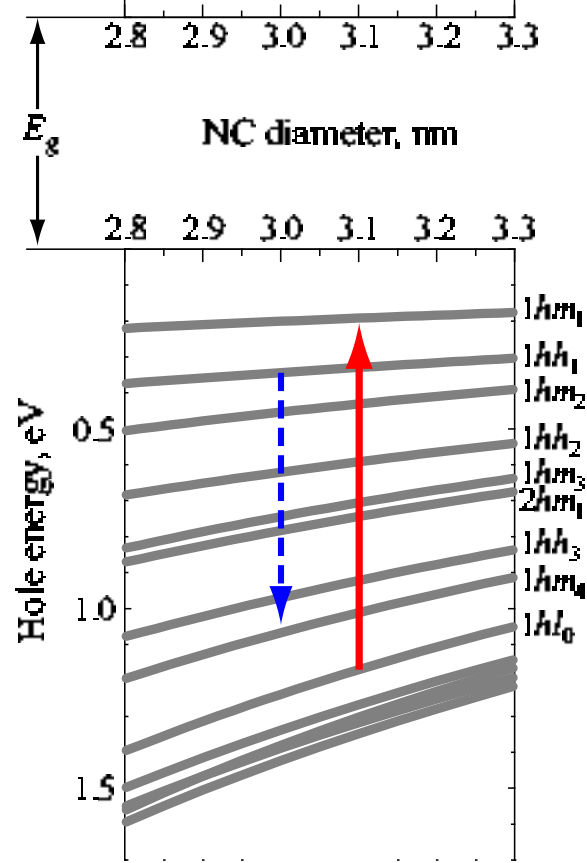
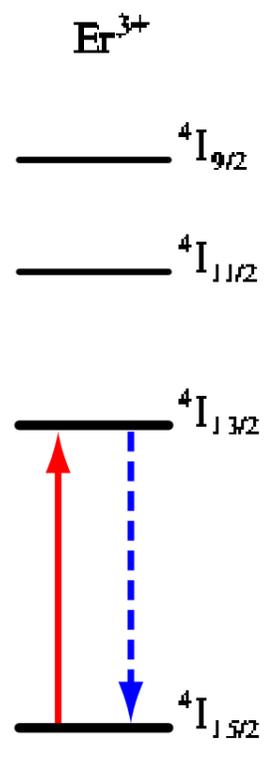
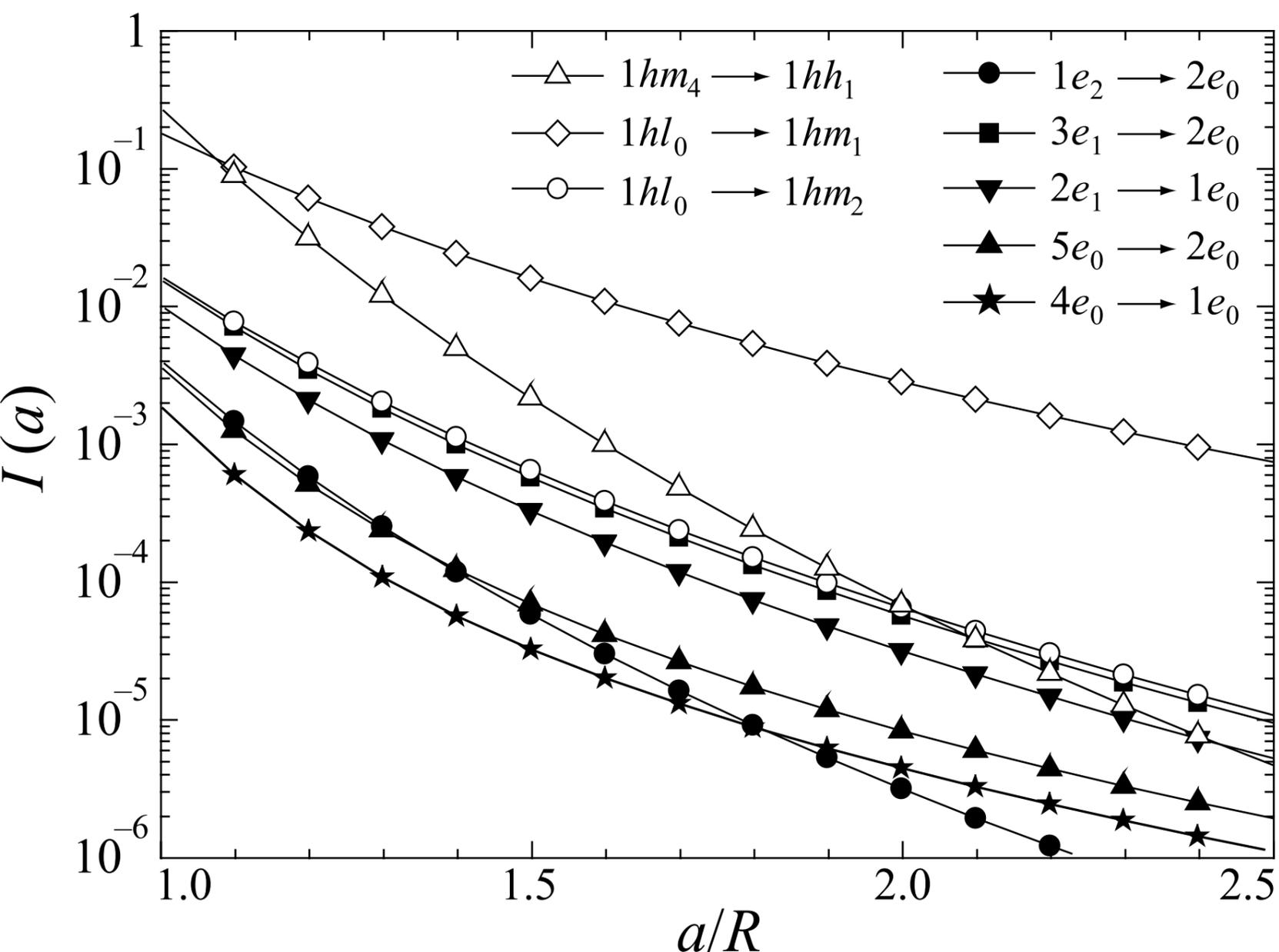
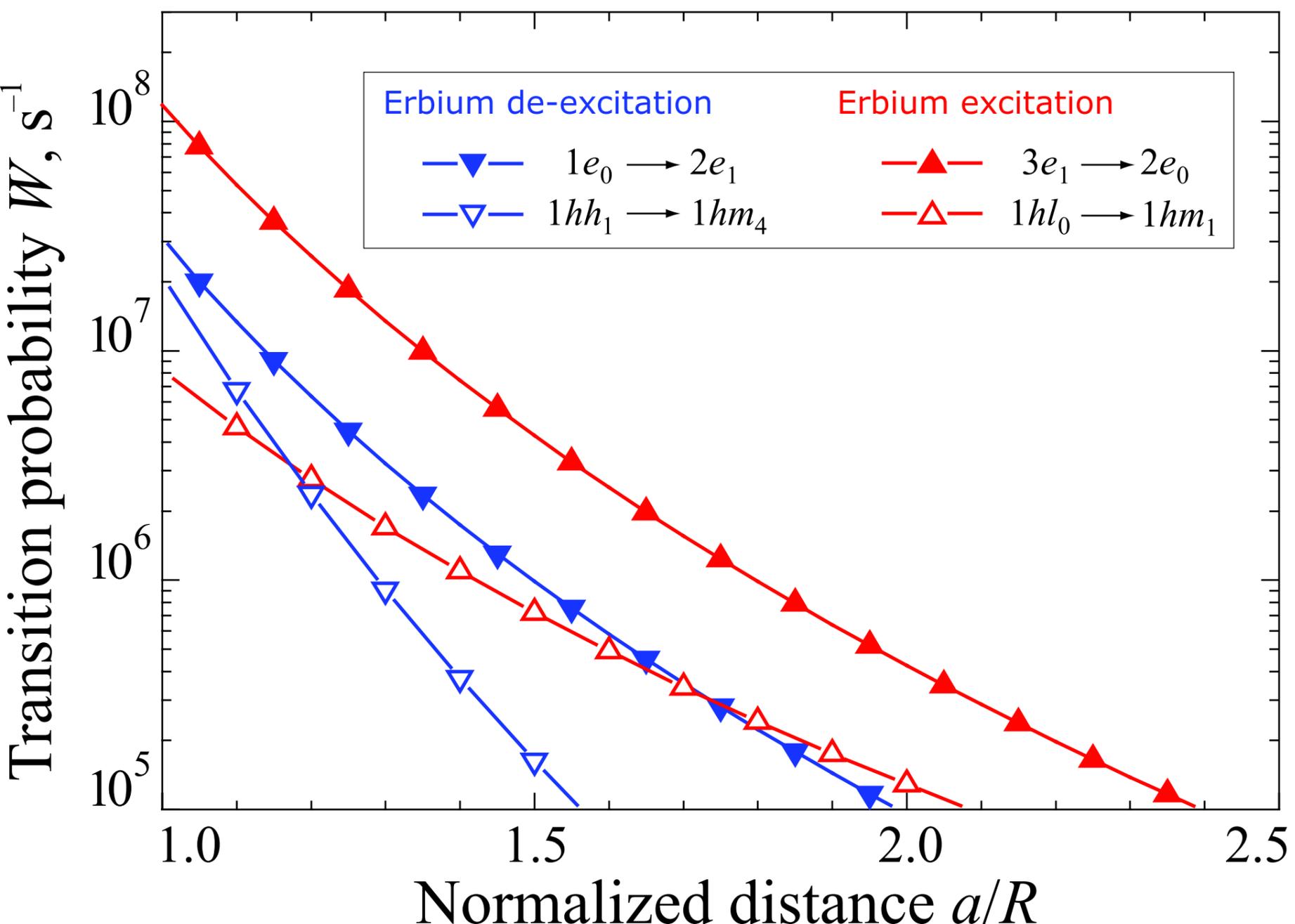
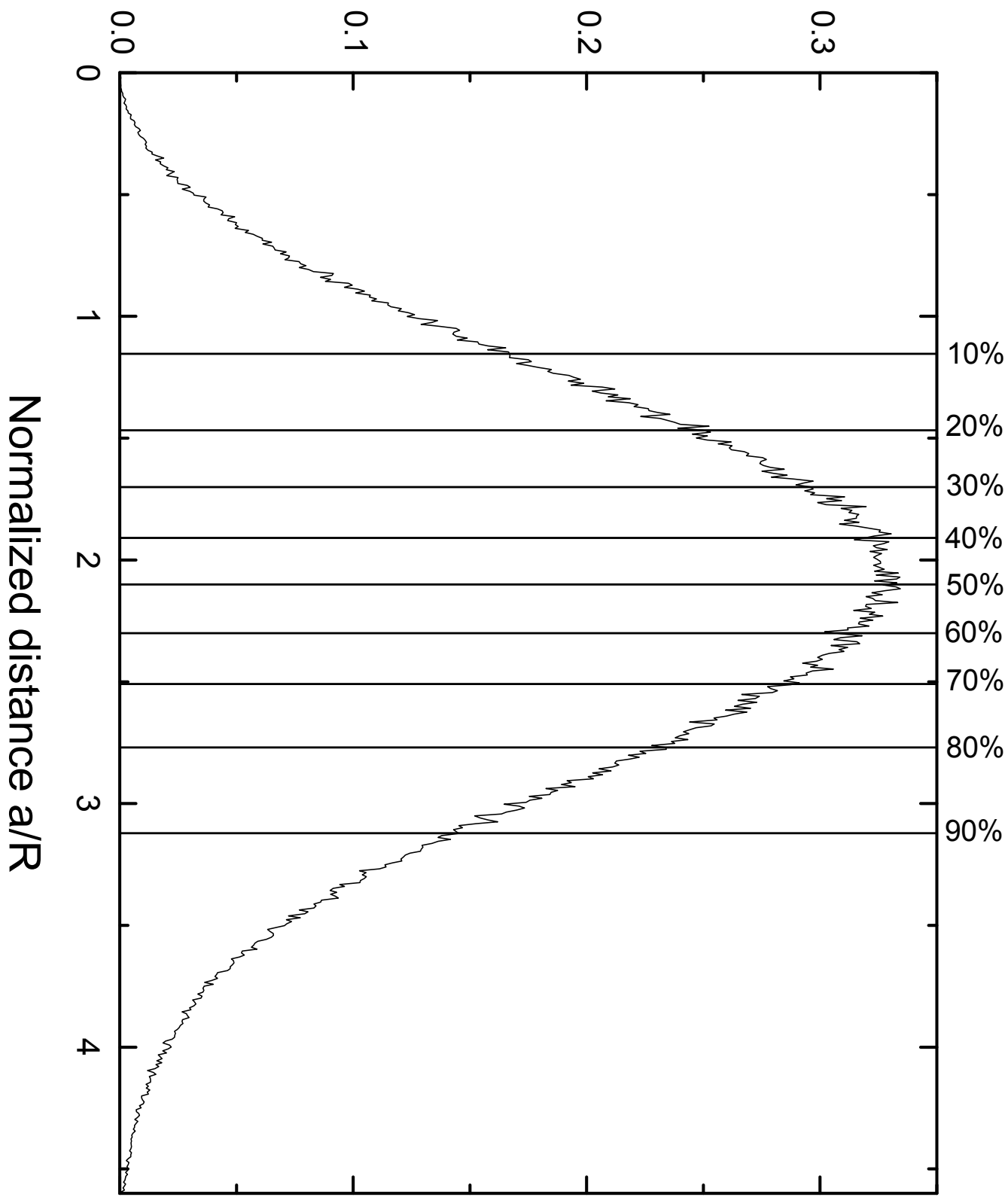
\end{document}